\documentclass[a4paper,twocolumn,10pt,accepted=2026-06-22,intlimits]{quantumarticle}
\pdfoutput=1

\usepackage[numbers,sort&compress]{natbib}

\usepackage{balance}
\usepackage{graphicx}%
\usepackage{tikz}
\usetikzlibrary{arrows}
\usetikzlibrary{decorations.pathreplacing}
\usetikzlibrary{decorations.pathmorphing}
\usetikzlibrary{3d, calc}

\usepackage{soul}

 \usepackage{xcolor}

\usepackage[breaklinks=true,colorlinks=true,linkcolor=blue,urlcolor=blue,citecolor=blue]{hyperref} %

\usepackage{amssymb}
\usepackage[intlimits]{amsmath} %
\usepackage{mathtools}
\allowdisplaybreaks %

\usepackage[capitalise]{cleveref}
\crefname{section}{Sec.}{Secs.}
\usepackage{physics}

\newcommand{\ii}{\mathrm{i}}
\newcommand{\ee}[1]{\mathrm{e}^{#1}} %

\newcommand{\integral}[3]{\int_{#2}^{#3} \!\! \mathrm{d} #1 \,}

\newcommand{\beq}{\begin{equation}}
\newcommand{\eeq}{\end{equation}}

\def\beqs#1\eeqs{%
  \begin{equation}
    \begin{split}
      #1
    \end{split}
  \end{equation}
}

\newcommand{\RR}{\mathbb{R}}
\newcommand{\MM}{\mathcal{M}}

\newcommand{\Orb}{\mathcal{O}_{\ket{\psi_i}}}

\newcommand{\ie}{{\it i.e.,}\ }

\begin{document}

\title{Geometric Speed Limit of State Preparation and Curved Control Spaces}%

\author{Maximilian Goll}
 \email{maximilian.goll@fysik.su.se}
 \affiliation{Freie Universität Berlin, Department of Physics, Arnimallee 14, 14195 Berlin}
 \affiliation{Nordita, Stockholm University and KTH Royal Institute of Technology, Hannes Alfvéns väg 12, SE-106 91 Stockholm, Sweden}%
\author{Robert H.~Jonsson}%
 \email{robert.jonsson@mau.se}
\affiliation{Nordita, Stockholm University and KTH Royal Institute of Technology, Hannes Alfvéns väg 12, SE-106 91 Stockholm, Sweden}
\affiliation{Department of Materials Science and Applied Mathematics, Malmö University, SE-205 06, Malmö, Sweden}

\begin{abstract}
The preparation of quantum many-body systems faces the difficulty that in a realistic scenario only few control parameters of the system may be accessible.
In this context, an interesting connection between the energy fluctuations during state preparation and its geometric length as measured by the  Fubini-Study metric was discussed in~\cite{bukov_geometric_2019}. 
An inspiring conjecture lower bounding the energy fluctuations by the minimal geometric length of all accessible state preparation protocols was put forward together with supporting examples and numerical evidence.
However, we here show that the conjecture does not hold but can be violated if
the accessible parameter space has extrinsic curvature, when embedded into the space of all dynamically accessible states.
We illustrate this by a number
of generic qubit, qutrit and harmonic oscillator systems.
\end{abstract}

\maketitle

\section{Introduction}

State preparation is a foundational task in many envisioned uses of quantum technology.
It is the problem of how to steer a system, starting from a given initial state, to a desired target state using a given set of controls.
Typically, the initial state is easy and robust to prepare, whereas the target state is valuable or useful for the application.
A paradigmatic example is adiabatic quantum computing~\cite{albash_adiabatic_2018} where the solution to hard optimization problems is encoded in the properties of a target state.
Given that state preparation can be used to solve hard problems, it is not surprising that 
state preparation itself can be difficult to achieve, depending on the system's details.

Ultimate limits to the feasibility of state preparation can be quantified in terms of quantum speed limits.
For example, the time $t_\perp$ in which a system can evolve from one state into an orthogonal state is bounded
by its energy variance $t_\perp \geq \tfrac{h}{4\Delta E}$ (Mandelstam-Tamm bound~\cite{mandelstam_uncertainty_1945}), and
its energy expectation value, $t_\perp\geq \tfrac{h}{4(E-E_0)}$ (Margolus-Levitin bound~\cite{margolus_maximum_1998}), where $E=\left<H\right>$ and $E_0$ is the Hamiltonian's groundstate energy.
Building on these fundamental works, today a range of results and extensions to various distinguishability measures, mixed states, open system dynamics and more are known~\cite{frey_quantum_2016}.
Geometric quantum speed limits go beyond the evolution time, which depends on the Hamiltonian's strength, and directly address the quantum geometric length of state paths.
In the context of state preparation, the work by Anandan and Aharonov~\cite{anandan_geometry_1990} is particularly interesting because it provided a first generalization to time-dependent Hamiltonians, and it introduced concepts from quantum geometry.
This avenue has proven fruitful and, for example, resulted in methods to systematically derive families of speed limits from families of distinguishability measures~\cite{pires_generalized_2016}.

In their most general form, quantum speed limits allow for all possible Hamiltonians to drive the evolution of the system.
For concrete applications such bounds often are  too weak to be relevant. 
Since in applications the set of available Hamiltonians is typically restricted, the time evolution may never saturate the most general speed limits, but much tighter limits may apply.
This applies in particular to state preparation problems and motivates the search for speed limits which take into account the restrictions of the accessible control space.
This approach was taken by~\cite{bukov_geometric_2019} which investigated the relation between the quantum geometry of the accessible Hamiltonians' ground state manifold and the average energy variance of the state path, as discussed in \cref{sec:setup}.
In particular, the work conjectured an intriguing geometric speed limit relating these two quantities. This conjecture, supported by analytical examples and numerical evidence, received significant attention and inspired follow-up research~\cite{lam_demonstration_2021,meinersen_quantum_2025}. %

In this article, we report that the conjecture does not hold.
Taking a differential geometry perspective, in \cref{sec:geometry}, we explain  problems of the conjecture which arise if the control space's ground state manifold has extrinsic curvature when embedded in the state of all controllable states, \ie states that can be reached dynamically.
\cref{sec:examples} illustrates this problem with generic counterexamples for qubit, qutrit and harmonic oscillator systems.
Several examples are of interest beyond the context of the above conjecture, as they achieve (unit fidelity) state preparation very close to, or even along the globally optimal path.

\section{Geometry of Accessible State Preparation }\label{sec:geometry}
\subsection{\label{sec:setup} State preparation and geometric distances}
We begin with a review of accessible state preparation and the connection between energetic and quantum geometric distance measures discussed in~\cite{bukov_geometric_2019}.
The setup is based on a system with a   family  of accessible Hamiltonians $\{\hat H(\lambda)\}$ with non-degenerate ground states. 
Here we assume that the accessible control space $\lambda\in\ \Lambda\subset\RR^p$  is given by  $p$ real-valued parameters, which may or may not be bounded. 
The problem at hand is to prepare a target state $\ket{\psi_*}=\ket{\psi_0(\lambda_*)}$, which is  the ground state of $\hat H(\lambda_*)$, starting from an initial state $\ket{\psi_i}=\ket{\psi_0(\lambda_i)}$,  the ground state of $\hat H(\lambda_i)$.
For this, one needs to find an appropriate time-dependent Hamiltonian, that is a path
$\lambda(t)$ %
such that the time evolution 
\beq
    \ket*{\dot\psi(t)}=\dv{\ket{\psi(t)}}{t} = - \ii \hat H(\lambda(t))\ket{\psi(t)},
\eeq
with initial condition $\ket{\psi(t=0)}=\ket{\psi_i}$, %
arrives at $\ket{\psi(T)}=\ket{\psi_0(\lambda_*)}$
after some time $T$.
It is natural, although not essential, to assume that $\lambda(0)=\lambda_i$ and $\lambda(T)=\lambda_*$.

The conjecture of~\cite{bukov_geometric_2019} compares two natural ways to measure the state preparation length: One uses a geometric notion of length for the Hamiltonian path $\lambda(t)$, %
the other uses an energetic measure along the state path $\ket{\psi(t)}$.
In the following, we review how both distance measures can be understood in terms of the Fubini-Study metric on the state space of the system.

To measure the geometric length of the Hamiltonian path $\lambda(t)$ %
we need to equip $\Lambda$ with a metric. This can be achieved by mapping every accessible Hamiltonian $H(\lambda)$ to its ground state $\ket{\psi_0(\lambda)}$, which  we assumed to be non-degenerate.
We obtain the ground state manifold
\begin{equation}
    \MM_0=\left\{ \ket{\psi_0(\lambda)}\middle|\lambda\in\Lambda\right\},
\end{equation}
as a submanifold of the system's state space. Note that this manifold may have  a lower dimension than the control space $\Lambda$.
The  Fubini-Study metric, which is the natural metric on state space, induces a metric on $\MM_0$, and thus on $\Lambda$, given by 
\beqs\label{eq:FubiniStudyLambda}
    ds^2 &= 1-|\braket{\psi_0(\lambda)}{\psi_0(\lambda+d\lambda)}|^2 
    \\&
    =\sum_{\mu\nu}g_{\mu\nu} d\lambda^\mu d\lambda^\nu+\mathcal{O}(d\lambda^3).
\eeqs
Here $g_{\mu\nu}=\chi_{[\mu\nu]}$ is the symmetric, and hence real, part of the quantum geometric tensor \cite{provost_riemannian_1980, kolodrubetz_geometry_2017, bengtsson_geometry_2006,hetenyi_fluctuations_2023}
\beq\label{eq:QGT}
    \chi_{\mu\nu}= \bra{\psi_0}\overset{\leftarrow}{\partial_\mu}\overset{\rightarrow}{\partial_\nu}\ket{\psi_0}
    -\bra{\psi_0}\overset{\leftarrow}{\partial_\mu}\ket{\psi_0}\bra{\psi_0}\overset{\rightarrow}{\partial_\nu}\ket{\psi_0}.
\eeq
With this metric, 
the geometric length of the Hamiltonian path $\lambda(t)$ is given by
\beqs\label{eq:gslength}
   l_g[\ket{\psi_0(\lambda(t))}]&= \integral{t}{0}{T}  \sqrt{g_{\mu\nu}\dv{\lambda^\mu}{t} \dv{\lambda^\nu}{t}} 
   \\& = \int_{\lambda(0)}^{\lambda(T)}  \sqrt{g_{\mu\nu}\dd \lambda^\mu\dd\lambda^\nu},
\eeqs
which is the length of the path $\ket{\psi_0(\lambda(t))}$ in the ground state manifold $\MM_0$ under the Fubini-Study metric. (See also \cref{fig:Manifold}.)

The second length measure, for the state path $\ket{\psi(t)}$, considers the time-dependent energy variance of the state under its evolution from $\ket{\psi_0(\lambda_i)}$ to $\ket{\psi_0(\lambda_*)}$ which is integrated up over time
$
    \integral{t}{0}T \sqrt{\delta E^2}.
$
At time $t$, the energy variance of the time-evolving state is
\beqs
    &\delta E^2(t)=\mel{\psi(t)}{\hat H^2(\lambda(t))}{\psi(t)}_c
    \\ &
    =\braket*{\dot\psi(t)}{\dot\psi(t)}-\braket*{\dot\psi(t)}{\psi(t)}\braket*{\psi(t)}{\dot\psi(t)}.
\eeqs

As shown in~\cite{anandan_geometry_1990}, and is evident from the previous equation,
this quantity is a function only of the state path $\ket{\psi(t)}$, but independent from the specific Hamiltonian used to generate the time evolution.
Furthermore, this quantity is identical to the length of the path $\ket{\psi(t)}$ %
as measured by the Fubini-Study metric~\cite{anandan_geometry_1990},
\beqs\label{eq:elength}
   l_E[\ket{\psi(t)}]&= \integral{t}{0}T \sqrt{\delta E^2}
   \\&=\integral{t}0T \sqrt{g\left( \dv{\ket{\psi(t)}}{t},\dv{\ket{\psi(t)}}{t}\right)}.
\eeqs

Thus we see that both the length of the Hamiltonian path $l_g$ and the length of the state path build on the Fubini-Study metric in state space. However, we highlight an important difference: The path $\ket{\psi_0(\lambda(t))}$, which yields the geometric length of the Hamiltonian path $l_g[\lambda(t)]$, remains in the ground state manifold $\MM_0$
of
accessible Hamiltonians.
In contrast, the state path $\ket{\psi(t)}$, which yields the energetic distance $l_E[\ket{\psi(t)}]$, will in general leave the ground state manifold $\MM_0$ and be driven into a larger subspace of the system's state space (as depicted in \cref{fig:Manifold}).

\begin{figure}
    \centering
    \begin{tikzpicture}[every node/.style={font=\fontsize{10pt}{12pt}\selectfont}]

\coordinate (A) at (-2,-1);
\coordinate (B) at (1.3, 0.5);
\coordinate (C) at (-0.2, 2.1);

\fill[gray!20] (-3,-2) -- (3,-1) -- (2,2) -- (-4,1) -- cycle;
\node[below right] at (-3.9,1) {$\mathcal{M}_0$};

\draw[red, thick, dashed]     
    (A) 
    .. controls (-1.5, -1) and (-1.2, 2.2) .. 
    (C)  %
    .. controls (0.3, 2.1) and (0.5, 0.3) .. 
    (B);

\node[below, red, thick, xshift=-8pt] at (-1.7,0.5) {$\ket{\psi(t)}$};

\draw[thick, blue] (A) .. controls (0,-1.25) and (0.3, -1) .. (B) node[midway, above, yshift=5pt, xshift=-10pt] {$\ket{\psi_0(\lambda(t))}$};

\fill[black] (A) circle (2pt) node[below] {$\ket{\psi_i}$};
\fill[black] (B) circle (2pt) node[below right] {$\ket{\psi_*}$};

\end{tikzpicture}
    \caption{ground state  manifold $\MM_0$ of the Hamiltonian family $\hat H(\lambda)$ as a submanifold in state space: The  target state $\ket{\psi_*}$ needs to be prepared starting from $\ket{\psi_i}$ under time evolution with some $\hat H(\lambda(t))$.
    The  Fubini-Study metric can measure both the length $l_g$ of the Hamiltonian ground state path $\ket{\psi_0(\lambda(t))}$, and $l_E$  of the state path $\ket{\psi(t)}$. Whereas the former lies inside $\MM_0$, the latter generally leaves $\MM_0$.
    }
    \label{fig:Manifold}
\end{figure}

\subsection{Conjectured distance inequality\label{subsec:inequality}}
The conjecture proposed in~\cite{bukov_geometric_2019} was that the length $l_E$ of a state path $\ket{\psi(t)}$ from $\ket{\psi_0(\lambda_i)}$ to $\ket{\psi_0(\lambda_*)}$, generated by a time-dependent accessible Hamiltonian $\hat H(\lambda(t))$ as described above, is always larger or equal to the length of the (shortest) geodesic from $\ket{\psi_0(\lambda_i)}$ to $\ket{\psi_0(\lambda_*)}$ inside the ground state manifold $\MM_0$:
\beq
    l_E[\ket{\psi(t)}]\geq d_{\MM_0}\left(\ket{\psi_i},\ket{\psi_*}\right)= \min_{\lambda(t)    }l_g[\ket{\psi_0(\lambda(t))}].
\eeq

In principle, any path $\lambda(t)$ %
connecting $\lambda_i$ to $\lambda_*$ in $\Lambda$ can be used to prepare the target state up to asymptotically vanishing corrections from the initial state, when traversed 
adiabatically, as asserted by adiabatic theorems~\cite{albash_adiabatic_2018}.
And therefore, as discussed in~\cite{bukov_geometric_2019}, the two length measures are equal, $l_E=l_g$, along adiabatic paths.
This is suggestive of the conjecture, since the (shortest) geodesic distance between initial and target state in $\MM_0$ is a lower bound to all other paths between them.

The central observation here is that the adiabatic state path $\ket{\psi(t)}$ does not leave the ground state manifold $\MM_0$.
Indeed, it is not difficult to see, that the conjecture always holds if a state path $\ket{\psi(t)}$ does not leave the ground state manifold $\MM_0$.
In this case, because $\ket{\psi(t)}\in \MM_0$, the Fubini-Study metric on the right hand side of \cref{eq:elength} is equal to the right-hand side of \cref{eq:gslength}. Hence for state paths in $\MM_0$, we have $l_E=l_g$, 
and the latter is larger or equal to the geodesic distance between initial and target state inside $\MM_0$. 

In general, however, the state path $\ket{\psi(t)}$ leaves the ground state manifold $\MM_0$.
The question therefore is, whether there can be state paths $\ket{\psi(t)}$ which are generated by Hamiltonians $\hat H(t)$ in the accessible control space, but nevertheless are shorter than the geodesic inside $\MM_0$, thus contradicting the conjecture.
Below we present a number of examples which answer this question in the affirmative.

\subsection{Extrinsic curvature of grounstate manifold}%

To understand how and when such counterexamples arise, a differential geometric perspective is helpful.
To this end we consider the set of all states which can be reached by accessible time-dependent Hamiltonians starting from the initial state $\ket{\psi_i}.$
We call this set the orbit of $\ket{\psi_i}$ under $\hat H(\Lambda)$ and denote it by $\Orb$.

The ground state manifold $\MM_0\subset\Orb$ is contained in the orbit, because we allow for time evolution of arbitrary duration and hence can move between states in $\MM_0$ adiabatically.
The orbit $\Orb$ itself can be viewed as a submanifold of the system's state space. 
That is we view $\MM_0$ as embedded in $\Orb$ which, in turn, is embedded in state space.
This allows us to equip $\Orb$ and $\MM_0$ with a metric by restricting the Fubini-Study metric from state space down to $\Orb$ and to $\MM_0$.

From this picture follows a sufficient condition for the original conjecture to hold:
If for any pair of points $(\ket{\psi_i},\ket{\psi_*})$ in $\MM_0$ their geodesic distance, \ie the length of the shortest geodesic connecting them, is the same in $\MM_0$ as in $\Orb$, then the conjecture holds in its original form.
Geometrically, for this to hold, one shortest geodesic joining the points $\Orb$  has to lie inside of $\MM_0$.\footnote{This condition is stronger than $\MM_0$ being a totally geodesic submanifold, which means that all geodesics in $\MM_0$ also are geodesics in $\Orb$. And it is weaker than $\MM_0$ being a totally convex submanifold, which means that all geodesics in $\Orb$ joining points in $\MM_0$ are contained in $\MM_0$ itself.}
On the other hand,  if $\MM_0$, as a submanifold of $\Orb$, has extrinsic curvature, then the geodesic joining $\ket{\psi_i}$ to $\ket{\psi_*}$ in $\Orb$, which appears in \cref{eq:modified_ineq}, may lie outside of $\MM_0$.
As a result $d_{\MM_0}(\ket{\psi_i},\ket{\psi_*})$ the geodesic distance in $\MM_0$ then is larger than $d_{\Orb}(\ket{\psi_i},\ket{\psi_*})$.

This opens for the possibility to violate the conjecture if there exists a schedule for the control Hamiltonian $\hat H(\lambda(t))$ that drives the state along a path close to the geodesic in $\Orb$.
Then $l_E[\ket{\psi(t)}]$ can get close to $d_{\Orb}(\ket{\psi_i},\ket{\psi_*})$ and shorter than $d_{\MM_0}(\ket{\psi_i},\ket{\psi_*})$. 
The following examples show that this can happen already in basic two-level and harmonic oscillator systems. 
They raise the question whether extrinsic curvature, in the above sense, is always sufficient for such shorter dynamical state paths to arise.

\section{Counterexamples}\label{sec:examples}

\begin{figure}[t]
    \centering
    \input{MaterialsPaper/qubit_sweep_theta0.25pi.pgf}
    \caption{Length comparison from the numerically solved ODEs for 49 evenly spaces values in $\phi_i\coloneq(0,2\pi)$ with $\theta=\pi/4$. The dynamical state path $l_E$ and the geodesic in the groundstate manifold $l_g$ are always lower-bounded by the global geodesic $\gamma$.
    }
    \label{fig:Sweep}
\end{figure}

In the following we discuss  the conjecture in the context of a two-level system (qubit), a linear shifted and a squeezed harmonic oscillator, and a three-level system (qutrit).
For all four we present cases which achieve unit fidelity state preparation while violating the conjecture.

The first example of a two-level system was considered in~\cite{bukov_geometric_2019}.
However, the configuration discussed there corresponds to the special case where $\MM_0$ is a flat submanifold of state space. We show that considering a generic curved accessible control space yields  counterexamples.
Secondly, we  consider a single harmonic oscillator with linearly shifted Hamiltonians as control space, which is easy to visualize since its metric corresponds to flat $\RR^2$.
Thirdly, we  consider a single harmonic oscillator with purely quadratic Hamiltonians as control space. 
Here the metric is more interesting and non-trivial,  illustrating that it may in general be difficult to decide whether the original conjecture applies to a given control space. 
Finally, the fourth qutrit example addresses linear control Hamiltonians, showing that the conjecture also does not generally apply in this important class of state preparation problems.

The contrast between the intriguing numerical evidence originally supporting the conjecture in~\cite{bukov_geometric_2019} and the counterexamples we give, raises the question which of the two cases is more generic. Is it more natural or common for the original conjecture to hold in a typical system of interest and counter-examples are fine-tuned instances? Or is the opposite more common?
An exhaustive treatment of this question goes beyond the scope of this work.
However,  we begin the section by presenting numerical evidence that suggests the existence of a large class  of counterexamples in two-level systems.

\subsection{Qubit with circular control space\label{subsec:qubit}}

\begin{figure}
    \centering
    \input{MaterialsPaper/qubit_bloch_theta0.25pi.pgf}
    \caption{Bloch sphere visualization of dynamical state path trajectories from the greedy algorithm for $\theta=\frac{\pi}{4}$, with $\phi_i=0.6\pi$ and $\phi_i'=1.4\pi$ (see \cref{subsec:qubit,fig:Sweep}).}
    \label{fig:greedy_bloch_sphere}
\end{figure}

We denote a general two-level (qubit) Hamiltonian  by
\beq
    \hat H(\theta, \phi)
    =-\frac{\omega}{2} \left(\sin\theta \left(\cos\phi\, \hat\sigma_x+%
    \sin\phi \,\hat\sigma_y  \right)+\cos\theta \, \hat\sigma_z \right),
\eeq
where
$\theta \in [0,\pi],\, \phi\in[0, 2\pi)$.
The sign convention is chosen such that in the Bloch sphere picture 
the same vector $\vec v=\left(\sin\theta \cos\phi\, ,\sin\theta \sin\phi \,, \cos\theta \,  \right)$ represents both the Hamiltonian, as  $\hat H(\theta, \phi)=-\frac\omega2 \, \vec v\cdot \hat{\vec\sigma}$, and its groundstate
\beq \label{eq:qubitstate}
\ket{\psi_0(\theta,\phi)}=\cos\frac{\theta}{2}\ket{0}+\sin\frac{\theta}{2}\ee{\ii\phi}\ket{1},
\eeq
as $\ketbra{\psi_0}{\psi_0}=\frac12 (\mathbb{I}+\vec v\cdot \hat{\vec\sigma} ).$ This yields a natural identification between accessible control Hamiltonians and their groundstate manifold.
In this standard notation, the Fubini-Study metric is
\beq
ds^2=\frac14 d\theta^2+\frac14\sin^2\theta d\phi^2,\label{eq:fs_qubit}
\eeq
\ie a quarter of the metric induced by the standard flat metric on $\RR^3$ on the Bloch sphere.

We now consider accessible control spaces given by circles on the Bloch sphere:
\beq
\Lambda_\theta = \left\{ (\theta,\phi) | \phi\in[0,2\pi) \right\}.\label{eq:Lambdatheta}
\eeq
For the special choice of $\theta=\tfrac\pi2$ the ground states of the control space form a great circle and hence a flat  ground state manifold $\MM_0$. 
Therefore, for $\Lambda_{\pi/2}$ the conjecture applies in its original form, as also verified analytically in~\cite{bukov_geometric_2019}.
However, for all other values of $\theta$ the ground state manifold is not a great circle and we will see that the original conjecture fails.

\subsubsection{Greedy algorithm exploration}
Before  constructing examples, which achieve state preparation with unit fidelity and violate the conjecture, we explore the existence of counterexamples numerically with a greedy control algorithm.
The idea of this algorithm is to always evolve the system to a given target state by driving the system as closely as possible along the geodesic from the current state $\ket{\psi(t)}$ to the target state.

To this end, the Hamiltonian $\hat H(t)$ is chosen from the accessible control Hamiltonians so as to maximize the overlap of the state's time derivative $\partial_t\ket{\psi(t)}= -\ii \hat H(t)\ket{\psi(t)}$ with the tangent vector of the geodesic from $\ket{\psi(t)}$ to the target state $\ket{\psi_*}$.
This means that the state's coordinates determine the Hamiltonian $\hat H(t)$ which, in turn, determines the time derivative of the state.
As detailed in \cref{app:greedyODE}, this leads to two coupled differential equations (ODEs) for the spherical coordinates of the state $\ket{\psi(t)}$.

We solve these ODEs numerically and compare, in \cref{fig:Sweep}, the length of the resulting dynamical state path $l_E$ to the length of the path in the groundstate manifold $l_g$ and the length of the global geodesic $\gamma$ connecting the initial to the target state.
\cref{fig:greedy_bloch_sphere} visualizes two examples of resulting state trajectories. 
Note that we end the numerical evolution of the ODEs, when the distance between dynamical state and target state reaches a small tolerance of $\Delta_{FS}=33\cdot 10^{-5}$ (corresponding to a fidelity of (below) $1-F\lesssim 1.1\cdot 10^{-7}$), because the numerical evaluation becomes challenging closer to the target state.
Hence, strictly speaking, we here do not demonstrate unit fidelity state preparation. 
However, the ODEs' numerical solutions and direction field suggest that the algorithm converges to unit fidelity on the scale of a few $\Delta_{FS}$. Hence we expect the numerical data in \cref{fig:Sweep} to indicate the existence of unit fidelity counterexamples.

The results in \cref{fig:Sweep} display an interesting but expected asymmetry.
Whereas $l_g$ and $\gamma$ are  symmetric around $\phi_*=\pi$, the algorithm yields slightly longer paths when starting from, for example, $\phi_i=3\pi/2$ than when starting from $\phi_i=\pi/2$.
The asymmetry  arises because the Hamiltonian $\hat H(\theta, \phi)=-\frac\omega2 \, \vec v\cdot \hat{\vec\sigma}$ generates a time evolution corresponding to clockwise rotation around $\vec v$ in the Bloch sphere picture.
This makes it easier to steer states starting from the right of the target ($\phi_i>0$) than from the left.

We see that for small positive values of $\phi_i$ the greedy algorithm is able to undercut $l_g$, thus violating the conjecture.
At a certain point  (here around $\phi_i=5\pi/4$) a crossover appears and the dynamic state paths become longer than $l_g$ for the remaining values of $\phi_i$.

From a topological point of view it is interesting to note that the shortest path in $\Lambda_\theta$ jumps from a clockwise orientation when $0<\phi_i<\pi$ to a counterclockwise orientation when $\pi<\phi_i<2\pi$.
This is in contrast to the azimuthal angles $\phi_H(t)$ which the greedy algorithm uses to control the qubit and which always lie between $\pi\leq \phi_H(t)\leq 2\pi$, although they may evolve non-monotonically.
In particular, for  $0<\phi_i<\pi/2$  where the algorithm is able to drive the state almost exactly along the geodesic, the time dependent Hamiltonian employed for this lies far away from the state path.

\subsubsection{Exact unit fidelity for opposite azimuth target state}
Therefore we now consider the configuration where the target state $\phi_*=\pi$ lies opposite to the initial state ($\phi_i=0$) on $\Lambda_\theta$.
Here we can state control schemes $\hat H(t)$ which achieve the state preparation with unit fidelity in finite time and along a path that is shorter than $l_g$ if $\theta\neq\pi/2$.

The simplest choice is to apply the constant Hamiltonian $\hat H(\theta,\pi/2)$. 
It is straightforward to calculate that this Hamiltonian evolves the qubit from $\ket{\psi_i}=\ket{\psi(\theta,0)}$ to $\ket{\psi_*}=\ket{\psi(\theta,\pi)}$ within time $T_\theta$ given by
\beq  T_\theta \,\omega = \sec ^{-1}\left(1-2 \csc ^2(\theta )\right).
\eeq
In the Bloch sphere picture, this evolution happens along a circle with radius
\beq
r_\theta=\frac{1}{\sqrt{2}}\left(\sin (\theta ) \sqrt{\cos (2 \theta )+3}\right),
\eeq
resulting in a dynamical state path length of
\beq l_E=\frac{r_\theta T_\theta \omega}2=\arcsin\left(\frac1{\sqrt{1+\cos^2\theta}}\right) \sqrt{1-\cos^4\theta}, \eeq
which is always less than the ground state manifold distance $l_g=\tfrac\pi2\sin\theta$.
For any value of $\theta\neq\tfrac\pi2$ there exists an interval around $\phi_*=\pi$ for which the constant Hamiltonian $\hat H(\theta,\phi_*/2)$ still results in $l_E<l_g$, \ie a violation of the conjecture.

Applying a constant Hamiltonian may appear nonphysical or impractical. However, it is not difficult to see, that continuous time-dependent choices of $\hat H(t)$ exist which begin with $\hat H(\theta,\phi_i)$ and end at $\hat H(\theta,\phi_*)$.
For example, for the case of opposite target and initial states,  $\phi_i=0$ and $\phi_*=\pi$, as just discussed, this can be achieved by a time-dependent control scheme $\hat H(\theta, \phi(t))$ where 
$\phi(t)=-\beta(t) \pi/2$ with
\beq\label{eq:betat}
\beta(t) = \begin{cases}
t/s, & 0\leq t \leq s \\
1, &  s < t \leq T + s \\
(t - T)/s, &  T + s < t \leq T + 2s
\end{cases}
\eeq
that is a function which linearly ramps up the Hamiltonian azimuthal angle in the time intervals $[0,s]$ and $[T+s,T+2s]$ while remaining constant for time $T$ in the middle of the scheme.
\Cref{fig:BlochSphere} shows a numerical example of such a solution for $\theta=\pi/4$. 

To see that these can achieve the state preparation with unit fidelity, note that for small values of $s$ the state $\ket{\psi(s)}$ is still close to the initial state with a small azimuthal angle $\phi(s)$.
The length of the middle time interval $T$ can then be chosen such that azimuthal angle of the state evolves   such that at $t=T+s$ the state has evolved to $\phi(T+s)=\pi-\phi(s)$. 
Due to the symmetry of the equations of motion, the interval $T+s<t<T+2s$ then is the mirror image of the first interval $0<t<s$ and the state ends up exactly at the target.

\begin{figure}[t]
    \centering
    \input{MaterialsPaper/BlochSphere.pgf}
    \caption{Bloch sphere representation of two-level system: If the accessible control space $\Lambda$ (solid, blue) does not correspond to a great circle on the Bloch sphere, then the original conjecture can be violated. For example, choosing numerical values $\omega=-1/2,s=4/10, T=1.4778$ %
    in \ref{eq:betat}, results in the (red, dashed) state path of length $l_E\approx0.81$, which is shorter than the distance inside $\Lambda=\pi/(2\sqrt2)\approx 1.11$. (Numerical simulation and visualization using~\cite{johansson_qutip_2013}.)}
    \label{fig:BlochSphere}
\end{figure}

\subsection{Harmonic oscillator with linear shifts}\label{subsec:coherentstates}

We consider a simple harmonic oscillator with linear shifts in its quadratures.
That is we use the two real control parameters $\lambda=(\lambda_q,\lambda_p)\in\RR^2$ and consider the Hamiltonian family
\beq\label{eq:HO-shift}
    \hat H(\lambda) = \frac\omega2 \left( (\hat q-\lambda_q)^2 +(\hat p-\lambda_p)^2\right).
\eeq
Using the displacement operator 
\beq 
\hat D_\lambda=\exp\left[\hat a^\dagger \left({\lambda_q+\ii \lambda_p}\right)/{\sqrt 2} -\hat a \left({\lambda_q-\ii \lambda_p}\right)/{\sqrt 2}\right]\label{eq:displop}
\eeq
we can write $\hat H(\lambda )=\hat D_\lambda \hat H(0,0)\hat D_\lambda^\dagger$. Hence the  ground state of $\hat H(\lambda)$ 
is the displaced vacuum
$%
\ket{\psi_0(\lambda)}=\hat D_\lambda\ket0.
$%

With respect to the two-dimensional control parameters, 
the Fubini-Study induces the metric
\beq 
ds^2=\frac12 d\lambda_q^2+\frac12 d\lambda_p^2.
\eeq
which shows that the induced metric on the control space %
is simply half of the standard metric on $\RR^2$. In particular, this means that geodesics in control space are given by straight lines. %

Under time evolution with  linearly shifted Hamiltonians,
displaced vacuum states evolve into other displaced vacuum states.
Thus, we can parametrize a state path
$
\ket{\psi(t)}=\ket{\psi_0(\mu(t))}
$
in terms of the displaced vacuum states it evolves through.
The time evolution induced by the (static) Hamiltonian $\hat H(\lambda)$ is visualized in $\RR^2$ by clockwise rotation of $\mu$ on a circle around $\lambda$.

\begin{figure}[t!]
    \centering
    \input{MaterialsPaper/PhaseSpaceCicleGeodesic.pgf}
    \caption{Harmonic oscillator with linear shifts and curved control space.
    The control space $\Lambda$ (blue, solid line) is a semi-circle (see \cref{eq:HO_linear_lambda}).
    The  Hamiltonian of \cref{eq:exact_parameter_HO} prepares the target state within time $0\leq t\leq \pi/\omega$, while driving the state $\ket{\psi(t)}$ (red, dashed line) along the geodesic in $\Orb$, thus violating the original version of the conjecture.}
    \label{fig:HO-linear-counterexample}
\end{figure}

A natural instance of the state preparation problem is now to pick $\lambda_i=(0,0)$ as the initial state, $\lambda^*=(q_0,0)$ as the target state and $\Lambda=[0,q_0]$ (or more precisely $\Lambda=\{(x,0)\in\RR^2| 0\leq x\leq q_0\}$) as the accessible control space.
Apart from the adiabatic solution, there also exist time-dependent Hamiltonians which accomplish the state preparation exactly and in finite time.
In fact, for any $\lambda'\in\RR^2$ it is possible to find a time-dependent Hamiltonian inside $\Lambda$ such that the system evolves from $\ket{\psi_0(\lambda_i)}$ to $\ket{\psi_0(\lambda')}$ in some finite time.
This means that the orbit $\Orb=\{\hat D_\lambda\ket0|\lambda\in\RR^2\}$ contains all displaced vacuum states.
Thus, for this $\Lambda$  the conjecture holds in its original form. 
Because inside $\Orb$, the minimal geodesic connecting $\ket{\psi_i}=\ket{0}$ to $\ket{\psi_*}=\ket{0,q_0}$ corresponds to the straight line between these points in $\RR^2$, which exactly corresponds to the accessible control space itself. 
Hence the conjecture holds with 
$ %
d_{\MM_0}\left((0,0),(q_0,0)\right)=q_0.
$%

However, curved accessible control spaces can violate the original version of the conjecture.
An analytically solvable counterexample is as follows.
For initial state $\lambda_i=(0,0)$ and target state $\lambda_*=(2,0)$ consider 
\beq\label{eq:HO_linear_lambda}
\Lambda=\left\{ (\lambda_q,\lambda_p)|0\leq \lambda_q\leq2,\lambda_p=\sqrt{1-(1-\lambda_q)^2}\right\}
\eeq
as the control space. It is depicted as the blue, solid semi-circle in \cref{fig:HO-linear-counterexample}.
The time-dependent Hamiltonian $\hat H(\lambda(t))$ with
\beq \label{eq:exact_parameter_HO}
\lambda_q(t) = 2 \sin^2(\omega t/2),\quad \lambda_p(t)= -\sin(\omega t), \quad 0\leq t\leq \pi
\eeq
achieves the state preparation while driving the state exactly along the geodesics
\beq
\ket{\psi(t)}=\ket{\psi_0( \mu(t) )},\quad \mu(t)=(\lambda_q(t),0)
\eeq
and arriving exactly at the target state at time $\ket{\psi(t=\pi)}=\ket{\psi(\lambda_q=2,\lambda_p=0)}$.
The length of the dynamical state path hence is $l_E=2$ whereas the length of the geodesic from $\lambda_i$ to $\lambda_*$ inside $\Lambda$ amounts to $d_{\MM_0}=\pi$.

This analytical example is only one instance of many choices of curved shapes of $\Lambda$ for which time-dependent Hamiltonians can be constructed that achieve the state preparation in finite time, while driving the state close to the geodesic in $\Orb$ thus violating the original version of the conjecture.

\subsection{Squeezed harmonic oscillator}\label{subsec:squeezedstates}
The previous examples have representations which are easy to visualize since the Fubini-Study metric corresponds to the flat Euclidean metric.
The following example has both a more complex representation and exhibits a non-trivial relation between the state path, the Hamiltonian path and the geodesic path connecting initial and target state.

We consider the family of Hamiltonians 
\beq\label{eq:squeezeHamiltonian}
\hat H(r,\theta)= \hat S_z \hat H_0 \hat S_z^\dagger
\eeq
obtained by squeezing the harmonic oscillator Hamiltonian
$%
\hat H_0=\frac\omega 2 \left(\hat q^2+\hat p^2\right)
$ %
with the operator
\beq
\hat S_z=\exp\left( \frac{z^*}2 \hat a^2-\frac{z}2\hat a^{\dagger 2}\right),\quad z=r\ee{\ii\theta},\,  r\geq0,\, \theta\in\RR.
\eeq
The ground state of $\hat H(r,\theta)$ is the squeezed vacuum %
\beq\label{eq:squeezedvacuumstate}
\ket{\psi_0(r,\theta)}=\hat S_z\ket0
=\!\sum_{n=0}^\infty \frac{(-\ee{\ii \theta}\tanh(r))^n}{\sqrt{\cosh(r)}} \frac{\sqrt{(2n)!}}{2^n n!}\ket{2n}.
\eeq
Similar to above, squeezed states evolve into squeezed states under arbitrary  Hamiltonians $\hat H(r,\theta)$ along orbits as depicted in  \cref{fig:EggsGeo}.
Note that the orbits are symmetric with respect to the axis through the origin and $(r,\theta)$. This will be used to construct counterexamples below.

The Fubini-Study metric induces the metric
\beq \label{eq:fs_squeezed}
ds^2=\frac12 dr^2+\frac18\sinh^2(2r)d\theta^2
\eeq
on the squeezed state parameters. It corresponds to the Poincaré disk~\cite{bachmann_dynamical_2017,chi_poincare_2024} which we obtain in its standard form under the coordinate transformation
$
x=\tanh(r)\cos\theta$ and $y=\tanh(r)\sin\theta,
$
after which
\beq
ds^2=\frac12\frac{\left(dx^2+dy^2\right)}{(1-x^2-y^2)^2}.
\eeq
From the Poincaré disk comparison we can deduce the shape of the geodesics in the squeezed states.
First, radial lines with constant $\theta$ (with a flip of $\theta\to\theta+\pi$ as they cross the origin) are geodesics.
Second, two points $(r,\theta_1)$ and $(r,\theta_2)$ with equal radial coordinate but different complex argument $\theta$ are connected by a geodesic that has a point of minimal radial coordinate $(r_{min},(\theta_1+\theta_2)/2)$ symmetrically in between the points.
As $r\to\infty$ the geodesic asymptotes to a straight radial line.

\begin{figure}[t!]
    \centering
    \input{MaterialsPaper/Eggs_and_Geodesics.pgf}
    \caption{Time evolution and geodesics of squeezed harmonic oscillator states:
    The dashed lines are orbits of time evolution under the static Hamiltonian $\hat H(r=2,\theta=0)$ in \cref{eq:squeezeHamiltonian}, whose ground state is marked by a black dot.
    The solid gray lines are geodesics under the Fubini-study metric \cref{eq:fs_squeezed}.
    }
    \label{fig:EggsGeo}
\end{figure}

For a control space with constant $\theta$-coordinate the original version of the conjecture applies because $\MM_0$ contains the shortest geodesic between any of its states.
However, a control space  with fixed $r$,
\beq
\Lambda_r=\left\{(r,\theta)|\theta\in[0,2\pi)\right\},
\eeq
can violate the original version of the conjecture in an interesting way:
Consider the state preparation task with initial state $\ket{\psi_i}=\ket{\psi_0(r,4\pi/3)}$ and target state $\ket{\psi_*}=\ket{\psi_0(r,2\pi/3)}$.
Time-dependent Hamiltonians of the form $\hat H(r,\theta(t))$ with $\theta(t)= (\beta(t)-1) 2\pi/3  $ and $\beta(t)$ as in \cref{eq:betat} can accomplish the state preparation while driving the state along a state path $\ket{\psi(t)}=\ket{\psi_0(\mu_r,\mu_\theta)}$ that is shorter than the geodesic inside $\Lambda_r.$
To this end, $s$ needs to be chosen short enough, such that $\mu_\theta(s)>\pi$. Then $T$ is chosen such that $\mu_\theta(s+T/2)=\pi$. 
With this choice, the symmetry of the Hamiltonian parameters about $t=s+T/2$, and the symmetry 
of the equations of motion, yield that the state evolves along %
\beq
\mu_r(2s+T-x) = \mu_r(x),\;\; \mu_\theta(2s+T-x)=2\pi-\mu_\theta(x).
\eeq
\Cref{fig:RTheta} shows a numerically constructed example of this kind.

\Cref{fig:RTheta} also highlights that the Hamiltonian ground state path, the geodesic path in $\MM_0$ and the state path are all different:
The Hamiltonian path $\ket{\psi_0(r,\theta(t))}$, that is the path of instantaneous ground states,
circles the ground state manifold counter-clockwise from $\theta=4\pi/3$ to $\theta=2\pi/3$. 
This path is thus twice as long as the geodesic connecting these states inside $\MM_0$ which runs clockwise and resulting in a geometric length of
\beq l_g=\frac\pi{24}\sinh^2(2r). \eeq
Nevertheless, the dynamic state path is much shorter than this geodesic.
This is because, as seen in \cref{fig:RTheta} $\mu_r\leq r$, such that the path remains inside the circle of $\MM_0$, where the  coefficient $\tfrac18 \sinh^2(2r)d\theta^2$ in the metric leads to a shorter energetic length $l_E<l_g$.

\begin{figure}[t!]
    \centering
    \input{MaterialsPaper/SqueezedECEeps0_003.pgf}
    \caption{State preparation for squeezed harmonic oscillator (\cref{subsec:squeezedstates}).
    The control space $\Lambda_{2}$ (solid, blue line) consists of all Hamiltonians with squeezing parameter $r=2$. 
    The initial state, at $\lambda_i$, has $\theta_i=-2\pi/3$, the target, $\lambda_*$, has $\theta_*=2\pi/3$.
    The  state path (red, dashed line) is obtained for numerical parameters $s=3\cdot10^{-3},\,\omega=2\pi, \,T=4.77\cdot10^{-6}$.
    Its length is $l_E\approx 42.24$ whereas the geodesic in $\Lambda_2$ measures $l_g\approx97.48$.
    }
    \label{fig:RTheta}
\end{figure}

\subsection{Linearly controlled qutrit}\label{sec:qutrit}
Linear controls are of particular interest, both from a practical and a theoretical perspective.
By this, we refer to Hamiltonians of the form $\hat H(\lambda)=\hat H_0+\lambda \hat K$ which are linear (or more strictly speaking affine) in the control parameter $\lambda$.
In fact, the analytic and numeric evidence presented in~\cite{bukov_geometric_2019} for the original conjecture, considered Hamiltonians only of this form, whereas the previous counterexamples do not fall into this class.
And from the discussion of the qubit example above, it follows that the conjecture holds for linearly controlled qubits.
However, in the following we present a counterexample for a qutrit with linear controls.

Specifically, we consider the Hamiltonian
\begin{equation}
    \hat H(\lambda)=\hat H_0+\lambda \hat K =\begin{pmatrix}-\omega&\ii\lambda &0\\-\ii\lambda &0&\ii\lambda a\\0&-\ii\lambda a&\omega  \end{pmatrix}, 
\end{equation}
with $\omega>0$ and the control problem in which the target state $\ket{\psi_*}=\ket{\psi_0(\lambda_*)}$ needs to be prepared starting from the initial state $\ket{\psi_i}=\ket{\psi_0(-\lambda_*)}$.
This is achieved, for all values of $\lambda_*$, by evolving the qutrit under the Hamiltonian $\hat H_0$ for a total time $T=\pi/\omega$.

To see this, note that the eigenstates of $\hat H(\lambda)$ are proportional to the three-vectors
\begin{equation}
    \ket{\psi_k(\lambda)}\propto \begin{pmatrix}
        a^2\lambda^2+\mu_k(\omega-\mu_k) \\ -\ii\lambda (\omega-\mu_k)\\ a\lambda^2
    \end{pmatrix}
\end{equation}
where $k=0,1,2$ and $\mu_k$ is the corresponding eigenvalue of $\hat H(\lambda)$. 
Hence one obtains $\ket{\psi_0(-\lambda)}$ from $\ket{\psi_0(\lambda)}$ by changing the sign of the second vector component, which is exactly achieved by $\mathrm{exp} (-\ii (\pi/\omega) \hat H_0)=\mathrm{diag}(1,-1,1).$
The Hamiltonian $\hat H(\lambda)$ remains gapped and no level-crossings occur as $\lambda$ changes, because its characteristic polynomial $\lambda^2 \omega - a^2 \lambda^2 \omega -\mu  (\lambda^2 + a^2 \lambda^2 + \omega^2) + \mu^3$ has a positive discriminant and, thus, three distinct real roots.

When $a\neq1$, the eigenvalues of $\hat H(\lambda)$, \ie the roots of the polynomial above are irrational.
Hence we use numerical calculations to obtain the length of the dynamical state path $l_E$ and the ground state manifold length $l_g$.
For the latter, we can use perturbation theory to obtain the tangent vector $\partial_\lambda\ket{\psi_0(\lambda)}$ to the ground state manifold  (by considering the effect of an infinitesimal change in $\lambda$ on the groundstate of $\hat H(\lambda)$).

As shown in \cref{fig:qutritnumerical}, we find that if $a>1$, there exists a critical value for $\tilde\lambda_a$ such that when $|\lambda_*|>\tilde\lambda_a$ we have $l_g>l_E$ thus disproving the original conjecture.

This numerical finding can be explained analytically by solving the case $a=1$ exactly, and then  considering perturbative corrections around $a=1+\epsilon$.
For $a=1$, the eigenvalues of $\hat H(\lambda)$ are $\{\pm\sqrt{\omega^2+2\lambda^2},0\}$.
Using the above expressions, one finds analytical expressions for $\ket{\psi_0(\lambda)}$ and $\partial_\lambda\ket{\psi_0(\lambda)}$. 
With these one finds the length of the path connecting the initial state $\ket{\psi_0(-\lambda_*)}$ to the target state $\ket{\psi_0(\lambda_*)}$ to be
\begin{equation}
    l_g=\sqrt2\arctan\left(\lambda_*\sqrt2/\omega\right)
\end{equation}
while the length of the dynamic path follows from the energy variance of $\ket{\psi_0(\lambda_*)}$ with respect to $\hat H(0)$ as
\beq
l_E=\frac{\pi  \lambda_*  \sqrt{\frac{\lambda_*^4+2 \lambda_*^2 \omega  \left(\sqrt{2 \lambda_*^2+\omega ^2}+2 \omega \right)+2
   \omega ^3 \left(\sqrt{2 \lambda_*^2+\omega ^2}+\omega \right)}{2 \lambda_*^2+\omega ^2}}}{\omega  \left(\sqrt{2
   \lambda_*^2+\omega ^2}+\omega \right)+\lambda_*^2}.
\eeq
Here the original conjecture holds, since for all finite values of $\lambda_*$ we have $l_g<l_E$ and in the limit of $\lambda_*\to\infty$, \ie when initial and target states are eigenstates of $\hat K$, we obtain
\beq
l_g^\infty=l^\infty_E=\pi/{\sqrt2}.
\eeq

If  we now expand $a=1+\epsilon$ we can calculate  the leading order perturbative corrections to the groundstate
$
    \ket{\psi_0(\lambda)}=\ket{\psi_0(\lambda)}_0+\epsilon \ket{\psi_0(\lambda)}_{1}+\mathcal{O}(\epsilon^2)
$
and to $\partial_\lambda\ket{\psi_0(\lambda)}$, accordingly.
From this we obtain leading order corrections to both $l_g$ and $l_E$.
In particular, we find that
\begin{equation}
    l_g^\infty=\pi/{\sqrt2}+5 \epsilon/({6\sqrt2})+\mathcal{O}(\epsilon^2),
\end{equation}
while the length of the dynamic state path $l_E^\infty=\pi/\sqrt2+\mathcal{O}(\lambda^2)$ between the eigenstates of $\hat K$ does not change to leading order. 
This explains the numerical observation presented in \cref{fig:qutritnumerical} of a critical value for $\lambda_*$ above of which the original conjecture is violated.

\begin{figure}
    \centering
    \input{MaterialsPaper/QutritLinearCE.pgf}
    \caption{Dynamical state path and groundstate manifold path length for the qutrit state preparation example of \cref{sec:qutrit}, with  $\omega=2$ and $a=1.5$. If $a>1$ there exists a critical value for $\lambda_*$ above of which $l_g>l_E$.}
    \label{fig:qutritnumerical}
\end{figure}

\section{Discussion and Conclusions}
In this work we have discussed the geometric speed limit for accessible quantum state preparation conjectured in~\cite{bukov_geometric_2019}.
We have derived conditions on the accessible control space under which the original conjecture holds, 
but we have also shown the existence of various, generic and instructive counterexamples to the conjecture.
We have explained the problems with the original conjecture from a differential geometric perspective attributing the origin of counterexamples to the extrinsic curvature of the system's groundstate manifold.

The counterexamples raise the question if the conjecture can be modified to still hold.
It follows from the counterexamples that instead of using the shortest path in the groundstate manifold, the  shortest path in the orbit $\Orb$, the set of all states that can be reached by accessible time-dependent Hamiltonians from the target state, needs to be considered.
Technically, this modification can be constructed as follows.
If  in the original conjecture the geodesic distance in $\MM_0$ is replaced by the geodesic distance in $\Orb$, which in general is shorter,
then the conjecture always holds:
\beq\label{eq:modified_ineq}
    l_E[\ket{\psi(t)}]\geq d_{\Orb}\left(\ket{\psi_i},\ket{\psi_*}\right).
\eeq
This is because now both the state path on the left hand side, and the geodesic path on the right hand side, lie within $\Orb$. Hence  $l_E$, which is the Fubini-Study length of $\ket{\psi(t)}$, cannot be shorter than the geodesic distance between initial and target state in $\Orb$.

This modified version of the conjecture is considerably weaker, of course. 
In fact, it is almost self-referential since it can be rephrased as the statement that the shortest dynamical state path from the initial to the target state is always shorter than or equal to any other such dynamical state path.
However, as the example in \cref{subsec:coherentstates} shows, the inequality in \cref{eq:modified_ineq} can be saturated, \ie there exist control spaces for which the system can be driven along the geodesic in $\Orb$ from initial state to target state, but the path inside $\MM_0$ is longer.
Hence, in order to derive stronger bounds for a state preparation problem, one will need to consider the specific properties of a  given accessible control space.

As we discussed  the original conjecture does apply if the ground state manifold contains a shortest geodesic between target and initial state within the controllable states.
In the absence of  topological obstructions this is the case when the ground state manifold is a flat submanifold of the controllable states.
In particular, this raises the question whether any non-zero extrinsic curvature of the accessible control space is sufficient for the existence of counter examples to the original conjecture.
Here our results from the greedy algorithm to the qubit example in  \cref{subsec:qubit} can be instructive for further research. 
For target states  to the right of the initial state ( positive $\phi_*-\phi_i$) the algorithm was not able to undercut the path length along the groundstate manifold for target states close to the initial state.
This is in line with the observation  in~\cite{bukov_geometric_2019} that the adiabatic path locally is a minimum of the path length.
However, for target states closely to the left of the initial state, the algorithm was able to drive the qubit closely to the global geodesic, thus undercutting the path along the groundstate manifold.
This is not contradicting the fact that the adiabatic path constitutes a local minimum of the path length, because the Hamiltonian path achieving it is far from the adiabatic path.

Hence, our results point towards a promising direction for future research which is to combine quantum control theory~\cite{glaser_training_2015,dalessandro_introduction_2021}, to describe the orbit of dynamically accessible states, with the differential geometry of the control space.
Studying the extrinsic curvature of the control space may allow for estimates of the difference between geodesics in control space and in the accessible states, leading to strong geometric speed limits for state preparation.

\section*{Acknowledgments}
We thank Tommaso Guaita and Jason Pye for inspiring discussions throughout this work. 
We thank Marín Bukov, Dries Sels and Anatoli Polkovnikov for their comments on the manuscript.
RHJ gratefully acknowledges support by the Wenner-Gren Foundations and the Wallenberg Initiative on Networks and Quantum Information (WINQ).
Nordita is supported in part by NordForsk.

\section*{Author Contributions}
The article presents the main results from the master research thesis of MG, which was supervised by RHJ. %
MG provided all code and figures for the article. 
RHJ constructed the greedy control algorithm.
LLMs were used for programming for syntax checks and translation between languages.

\appendix
\section{Fubini-Study Metric for Examples}
Here we present the calculations of the quantum geometric tensors
on the groundstate manifolds of the harmonic oscillator and qubit counterexamples in \cref{sec:examples}.

For the harmonic oscillator  with linear shifts in \cref{subsec:coherentstates}, the displacement operator in \cref{eq:displop} can be expressed via the Kermack–McCrea identity~\cite{kermack_professor_1931}
\beq
\hat D_\lambda=\ee{-\frac{\lambda_q^{2}+\lambda_p^2}{4}}\ee{\frac{\lambda_q+\ii\lambda_p}{\sqrt{2}}\hat a^{\dagger}}\ee{-\frac{\lambda_q-\ii\lambda_p}{\sqrt{2}}\hat a}.
\eeq
This makes it easy to calculate the derivatives of the coherent states $\ket{\psi(\lambda)}=\hat D_\lambda\ket{0}$ with respect to the components of $\lambda$ by differentiating the unitary operators. By commutating everything to the right of the displacement operator after using the sum rule for differentiation we obtain
\beqs\label{eq:dlam1}
\partial_{\lambda_q}\hat D_\lambda=&\frac{1}{\sqrt{2}}\hat D_\lambda\left(\hat a^\dagger-\hat a-\frac{\ii\lambda_p}{\sqrt{2}}\right)\\
\partial_{\lambda_p}\hat D_\lambda=&\frac{\ii}{\sqrt{2}}\hat D_\lambda\left(\hat a^\dagger+\hat a+\frac{\lambda_q}{\sqrt{2}}\right).
\eeqs
Plugging these equations into \cref{eq:QGT} the displacement operator and its adjoint cancel to the identity and only the action of $\hat a$ and $\hat a^\dagger$ on the vacuum has to be considered. This yields the metric
\beq
\chi=\frac{1}{2}
\begin{bmatrix}
   1 & \ii\\
   -\ii & 1
\end{bmatrix} 
\quad\Rightarrow\quad g=\frac{1}{2}\mathbb{I}
\eeq

For the qubit in \cref{subsec:qubit}, we consider the state as in \cref{eq:qubitstate}, for which %
\beqs
\partial_\theta\ket{\psi(\theta,\phi)}=&-\frac{1}{2}\sin{\frac{\theta}{2}}\ket{0}+\frac{1}{2}\cos{\frac{\theta}{2}}\ee{\ii\phi}\ket{1}\\
\partial_\phi\ket{\psi(\theta,\phi)}=&\ii\sin{\frac{\theta}{2}}\ee{\ii\phi}\ket{1}.
\eeqs
Substituting these into \cref{eq:QGT}   gives 
\beq
\chi=\frac{1}{4}
\begin{bmatrix}
    1 & \ii\sin{\theta}\\
    -\ii\sin{\theta} & \sin^2\theta
\end{bmatrix}
\quad\Rightarrow\quad g=\frac{1}{4}
\begin{bmatrix}
    1 & 0\\
    0 & \sin^2\theta
\end{bmatrix}.
\eeq

For the squeezed harmonic oscillator in \cref{subsec:squeezedstates}, the derivation of the QGT is more involved.
The state \cref{eq:squeezedvacuumstate} can be expressed in the Fock basis as
\beqs
    &\ket{\psi(r,\theta)}
    =\frac1{\sqrt{\cosh(r)}}\sum_{n=0}^\infty (-\ee{\ii \theta}\tanh(r))^n \frac{\sqrt{(2n)!}}{2^n n!}\ket{2n}
    \\
    &\quad=\sum_{n=0}^\infty  \frac{\left(-\frac{1}{2}\right)^n \ee{\ii \theta  n} \tanh ^n(r) \sqrt{(2 n)! \text{sech}(r)}}{n!} \ket{2n}.
\eeqs
The derivatives can be computed and simplified to
\beqs
    &\partial_r \ket{\psi(r,\theta)}
    =\sum_{n=0}^\infty  \frac{2^{-n-1} \sqrt{(2 n)!} \sinh (r)}{n!
   \cosh ^{\frac{3}{2}}(r)}\;\\
   &\phantom{\partial_r \ket{\psi(r,\theta)}
    =} \times\left(2 n  \text{csch}^2(r)-1\right) \left(-\ee{\ii \theta } \tanh (r)\right)^n\ket{2n}
\eeqs
and
\beqs
    &\partial_\theta \ket{\psi(r,\theta)}
   =\sum_{n=0}^\infty  \frac{\ii \left(-\frac{1}{2}\right)^n n \ee{\ii \theta  n} \tanh ^n(r)}{n!}\;\\
   &\phantom{\partial_\theta \ket{\psi(r,\theta)}=} \times \sqrt{(2 n)! \text{sech}(r)}\ket{2n}.
\eeqs
The relevant overlaps are
\begin{align}
   &\bra{\psi(r,\theta)}\overset{\leftarrow}{\partial}_\theta\overset{\rightarrow}{\partial}_\theta \ket{\psi(r,\theta)} =  \sinh ^2(r) (3 \cosh (2 r)+1)/8 \nonumber\\
&    \bra{\psi(r,\theta)} \overset{\rightarrow}{\partial}_\theta \ket{\psi(r,\theta)} = \ii \sinh ^2(r)/2\nonumber\\
 &    \bra{\psi(r,\theta)}\overset{\leftarrow}{\partial}_r  \overset{\rightarrow}{\partial}_r \ket{\psi(r,\theta)}  = 1/2\nonumber\\
  &  \bra{\psi(r,\theta)}\overset{\rightarrow}{\partial}_r \ket{\psi(r,\theta)} =0\nonumber\\
   & \bra{\psi(r,\theta)}\overset{\leftarrow}{\partial}_\theta \overset{\rightarrow}{\partial}_r \ket{\psi(r,\theta)} = -  \ii \sinh (2 r)/4.
\end{align}
Hence,  the QGT and thus the Fubini-Study metric take the form
\beqs
\chi&=
\begin{bmatrix}
    \frac{1}{2} & \frac{\ii}{4}\sinh(2r)\\
    -\frac{\ii}{4}\sinh(2r) & \frac{1}{8}\sinh^2(2r)
\end{bmatrix}
\\&\quad
\Rightarrow
g=
\begin{bmatrix}
    \frac12 &0\\
    0& \frac18\sinh^2(2r)
\end{bmatrix}.
\eeqs

\section{Greedy Continuous Control Implementation
\label{app:greedyODE}}
Here we give details on the greedy control scheme of \cref{subsec:qubit}.
Its idea is to choose the control Hamiltonian such that it always drives the state as closely as possible along the direction of the geodesic between the current state and the target state.

The density matrix of a qubit can be written as
\beq
\hat \rho=\frac12\left(\mathbb{I}+\vec \rho\cdot \hat{\vec \sigma}\right)
\eeq
where $\vec \rho$ is the vector representing the state on the Bloch sphere, and $\hat{\vec \sigma}=(\hat\sigma_x,\hat\sigma_y,\hat\sigma_z)$ is a three vector of Pauli matrices.
If the qubit state is pure, as in our case, the vector has unit length $|\vec \rho|=1$.
Similarly, we write the Hamiltonian acting on the qubit as
\beq
\hat H=-\frac\omega2 \vec h\cdot\hat{\vec \sigma}.
\eeq
The Schrödinger equation then yields
\beq
\partial_t \hat \rho =-\ii\comm{\hat H}{\hat \rho}=-\frac\omega2 \vec h\times\vec \rho
\label{eq:blochschrodinger}
\eeq
which implies that the time derivative of the state in Bloch sphere picture is represented by
$
\dot{\vec \rho}=- \omega\vec h\times\vec \rho.
$

The idea of the algorithm is to choose the Hamiltonian such that $\dot{\vec \rho}$ aligns as closely as possible with the geodesic between the current state and the target state $\hat\rho_*=\frac12\left(\mathbb{I}+\vec\rho_*\cdot\hat{\vec \sigma}\right)$.
A tangent vector of this geodesic is given by
$
\left(\vec \rho\times\vec \rho_*\right)\times\vec \rho = \vec \rho_*-\left(\vec \rho\cdot\vec \rho_*\right)\vec \rho.
$
Hence, $\vec h$ needs to maximize the scalar product between the states' time derivative and the desired tangent vector:
\beqs 
&\dot{\vec \rho}\cdot\left( \vec \rho_*-\left(\vec \rho\cdot\vec \rho_*\right)\vec \rho\right)
= -\frac\omega2(\vec h\times \vec\rho) \cdot  \vec \rho_*
=\frac\omega2  \vec h\cdot (\vec \rho_*\times\vec \rho).
\eeqs
With the following spherical coordinates for the state Bloch vectors
$\vec h=(\cos\phi\sin\theta,\sin\phi\sin\theta,\cos\theta)$,
$\vec \rho_*=(\sin\theta,0,\cos\theta)$, and
$\vec\rho=(\cos\chi\sin\eta,\sin\chi\sin\eta,\cos\eta)$,
we have 
\beq
\vec \rho_*\times\vec\rho
=\begin{pmatrix}
    -\sin \eta  \cos \theta \sin \chi  \\
    \sin \eta  \cos \theta  \cos \chi -\cos \eta  \sin \theta \\
    \sin \eta \sin\theta  \sin\chi
\end{pmatrix}.
\eeq
Since the polar angle $\theta$ is fixed (recall that we are working with the set $\Lambda_\theta$ of \cref{eq:Lambdatheta} as accessible control Hamiltonians), we can only optimize the Hamiltonian's azimuthal angle $\phi$. 
Hence, to maximize %
$\vec h\cdot(\vec \rho_*\times\vec \rho)$, we choose $\phi$ such that
\beqs 
&\cos\phi=-\frac{\cos\theta\sin\eta\sin\chi}{\|\vec \rho_*\times\vec \rho\|_{1,2}},
\\ &
\sin\phi=\frac{\sin \eta  \cos \theta  \cos \chi -\cos \eta  \sin \theta }{\|\vec \rho_*\times\vec \rho\|_{1,2}},
\eeqs
where  the norm of the first two elements of $\vec \rho_*\times\vec\rho$ is
\beqs
&\|\vec \rho_*\times\vec \rho\|_{1,2} =
\\ &\sqrt{\left(\sin \eta  \cos \theta \sin \chi \right)^2 +\left(\sin \eta  \cos \theta  \cos \chi -\cos \eta  \sin \theta\right)^2 }\label{eq:optHphi}.
\eeqs

Inserting the optimized choice for $\phi$ into the Schrödinger equation \cref{eq:blochschrodinger} results in the ODEs
\beqs
\dv\eta{t}&=\frac{\sqrt{2} \omega  \sin \eta  \sin ^2\theta  (\cot \eta  \cos \chi -\cot \theta )}{\sqrt{1-\sin (2 \eta ) \sin (2 \theta ) \cos \chi -\cos (2 \eta ) \cos(2 \theta )}},
\\
\dv\chi{t}
&=-\omega  \cos \theta
\\&\quad
-\frac{\sqrt{2} \omega  \cos \eta  \cot \eta  \sin ^2\theta  \sin \chi }{\sqrt{1-\sin (2 \eta ) \sin (2 \theta ) \cos \chi -\cos (2 \eta ) \cos (2 \theta
   )}} .
\eeqs
Numerical solutions of these equations show that the resulting state paths can take very long time to converge to the target state.
In fact, it is conceivable that the target state is only reached asymptotically for infinite times.

To circumvent this difficulty, it is helpful to parametrize the dynamical state path in terms of its arc length.
The time derivative of the arc length (with respect to the Fubini study metric \cref{eq:fs_qubit}) is
\beq
\dv{s}{t} = \frac12\sqrt{\sin^2\eta \left(\dv\chi{t}\right)^2+\left(\dv{\eta}{t}\right)^2}.
\eeq
Then with $\dv{\eta}{s}=\dv{\eta}{t}\left( \dv{s}{t}\right)^{-1} $ we obtain
\begin{widetext}
\beqs
&\dv{\eta}{s} %
=\frac{2 (\cot \eta  \cos \chi -\cot \theta )}{\sqrt{(\cot\theta -\cot \eta \cos\chi)^2+\left(\sqrt{\cot ^2\theta \left( \sin ^2\eta \csc ^2\theta+\cos (2 \eta )-\sin (2 \eta ) \cot\theta   \cos\chi\right)}+\cos\eta \cot\eta  \sin \chi\right)^2}},
\\
&\dv{\chi}{s}
=\frac{-2 \csc\eta \left(\sqrt{\cot^2\theta \left( \sin^2\eta \csc^2\theta +\cos(2 \eta )-\sin (2 \eta ) \cot\theta  \cos\chi\right)}+\cos\eta   \cot\eta \sin\chi\right)}{\sqrt{(\cot\theta -\cot\eta  \cos\chi)^2+\left(\sqrt{\cot ^2\theta \left(\sin ^2\eta \csc ^2\theta+\cos (2 \eta )-\sin (2 \eta ) \cot\theta\cos\chi\right)}+\cos\eta  \cot\eta  \sin\chi \right)^2}}.
\eeqs
\end{widetext}

The numerical integration of these equations becomes challenging as the state approaches the target state.
To save resources, we therefore abort the integration when the distance between state and target falls below a certain tolerance.
To interpret this distance in terms of fidelity between the state and the target state, note that the Fubini-Study distance between two pure states $d_{FS}=\gamma/2$ equals half the opening angle $\gamma$ between their Bloch vectors.
The fidelity between two states is
\beq
F %
=\frac{1+\cos\gamma}2
=\frac{1+\cos(2 d_{FS})}2\approx 1-d_{FS}^2
\eeq
hence the tolerance of $d_{FS}<\Delta_{FS}=33\cdot 10^{-5}$, which we chose for the numerical results presented above, corresponds to a fidelity of $1-F\lesssim 1.1\cdot 10^{-7}$.

Going beyond the scope of this work, an investigation of the analytical properties of the ODE for points close to the target state may allow to show whether the numerical solutions end up reaching the target with unit fidelity.
The behavior of the numerical solutions we obtain here suggests that the state paths would have to be continued by a length less than $2\Delta_{FS}$ in order to reach the target state with unit fidelity, because the absolute distance to the target tends to drop at a high rate as the state approaches the target.

\bibliographystyle{plainnat}
\bibliography{article_geo_limit_refs}%

\end{document}